\documentclass[12pt]{article}

\usepackage{amsfonts}
\usepackage{amsmath}
\usepackage{epsfig}
\usepackage{latexsym}
\usepackage{graphicx}
\usepackage{subfigure}
\usepackage{amssymb}
\numberwithin{equation}{section}
\allowdisplaybreaks

\def\spa#1{\phantom{\fbox{\rule[-#1cm]{0cm}{0cm}}}}

\def\be{\begin{equation}}
\def\ee{\end{equation}}
\def\bea{\begin{eqnarray}}
\def\eea{\end{eqnarray}}

\renewcommand{\thefootnote}{\fnsymbol{footnote}}

\begin{document}

\hfuzz=100pt
\title{
\begin{flushright} 
\small{WITS-CTP-126}, \small{YITP-14-1}
\end{flushright} 
{\Large \bf{Interpreting canonical tensor model in minisuperspace}}}
\date{}
\author{Naoki Sasakura$^a$\footnote{sasakura@yukawa.kyoto-u.ac.jp} and Yuki Sato$^{b}$\footnote{Yuki.Sato@wits.ac.za}
  \spa{0.5} \\
\\
$^{a}${\small{\it Yukawa Institute for Theoretical Physics, Kyoto University,}}
\\ {\small{\it Kyoto 606-8502, Japan}}\\
\\
$^{b}${\small{\it National Institute for Theoretical Physics, }}
\\ {\small{\it Department of Physics and Centre for Theoretical Physics,}}
\\ {\small{\it University of the Witwartersrand, WITS 2050,}} 
\\ {\small{\it South Africa}}
  \spa{0.3} 
}
\date{}

\maketitle
\centerline{}

\begin{abstract}
Canonical tensor model is a theory of dynamical fuzzy spaces in arbitrary space-time dimensions.
Examining its simplest case, we find a connection to a minisuperspace model of general relativity in arbitrary dimensions.
This is a first step in interpreting variables in canonical tensor model based on the known language of general relativity.   
\end{abstract}

\renewcommand{\thefootnote}{\arabic{footnote}}
\setcounter{footnote}{0}

\newpage
\section{Introduction}\label{intro}
Locality is an important issue in quantum gravity: due to the diffeomorphism invariance, one cannot define any well-defined local observables in general space-time,
\footnote{In the AdS/CFT correspondence \cite{Maldacena:1997re}, such local observables can be defined on an infinite conformal boundary in an Anti-deSitter space.}
which is also closely related to Bekenstein's entropy bounds \cite{Bekenstein:1981, Bekenstein:2004sh}.
This notion might suggest that space-time has a fundamental fuzzyness and eventually aquires a smooth manifold structure at a long-distance scale.
Since fuzzy spaces posess no notion of dimensionality and locality in general, 
their space-time dimensions is not a parameter but something ought to be determined through dynamics; 
additionally, as a result of dynamics, locality has to be favoured at a long-distance scale.
   
\textit{Canonical tensor model} is one of such trials introduced by one of the current authors
as a theory of dynamical fuzzy spaces \cite{Sasakura:2011sq, Sasakura:2012fb, Sasakura:2013gxg, Sasakura:2013wza}.
The fundamental variable is a rank-three tensor, which specifies the structure of fuzzy spaces; 
the time evolution of fuzzy spaces can be determined by a Hamiltonian flow.
Somewhat amazingly, the Hamiltonian can be uniquely fixed under some reasonable assumptions \cite{Sasakura:2012fb}.
So far, it has been shown that locality is favoured as a result of dynamics at least when indices can take only two 
values \cite{Sasakura:2013wza}.

The canonical tensor model may not be an isolated model and is expected to be related with other types of tensor models.
Here taking an overlook at history of tensor models, let us introduce several cousins.
Dating back to the original introduction of tensor models \cite{ambjorn, sasakura, godfrey},
the first motivation was to construct a model of higer-dimensional simplicial quantum gravity
as a natural extension of matrix models which describe two dimensional simplicial quantum gravity.
As far as a symmetric tensor is concerned, 
this program did not work \cite{sasakura, DePietri:2000ii}. 
However, tensor models with unsymmetric tensors called \textit{coloured tensor models} have been proposed \cite{Gurau:2009tw};
the newly introduced ``colour'' degrees of freedom turn out to fit together well with simplicial geometries and this line of works are still in progress \cite{Gurau:2011xp}.
On the other hand, apart from the interpretation as a simplicial quantum gravity, 
tensor models have developed into so-called \textit{group field theories} 
letting indices group-valued \cite{Boulatov:1992vp, Ooguri:1992eb, DePietri:1999bx, Freidel:2005qe, Oriti:2011jm} and
the canonical tensor model as a theory of dynamical fuzzy spaces which we argue in this paper. 

Most importantly, as a theory of quantum gravity the canonical tensor model ought to be related to general relativity in arbitrary dimensions as well.
Therefore, interpreting rank-three tensors in the canonical tensor model based on the established language of general relativity is 
absolutely imperative. However, this part is still veiled in mystery.
The purpose of this paper is to make progress in that direction:
in a simple situation that indices can take only a single value ($N=1$), 
we have identified the tensors as variables of general relativity in minisuperspace.
The paper is organised as follows:
in Section \ref{minimal}, we examine the canonical tensor model with $N=1$ and 
derive an effective action written by its degrees of freedom.
In Section \ref{mini}, we consider the Einstein-Hildert action in arbitrary dimensions and 
reduce it by the minisuperspace ansatz.
As a result, we obtain a corresponding effective action, which is nothing but the effective action derived from the canonical tensor model.
In Section \ref{dis}, we summarise our results.

\section{Canonical tensor model}\label{minimal}
The canonical tensor model has been developed by a series of works \cite{Sasakura:2011sq, Sasakura:2012fb, Sasakura:2013gxg, Sasakura:2013wza}, 
and designed to describe a theory of dynamical fuzzy space in the canonical formalism. 
Since the fuzzy space itself does not necessarly include information of space dimensions a priori,
the canonical tensor model might have a potential to describe quantum gravity in arbitrary dimensions. 
Physical degrees of freedom in the canonical tensor model are a rank-three tensor $M_{abc}$ and its canonical conjugate $P_{abc}$; 
its classical and even quantum dynamics can be completely determined by the unique Hamiltonian in principle.  
However, so far, the role of the rank-three tensors is unclear: 
we don't know how to interpret them in the standard language of gravitation. 
The main purpose of this paper is to address this issue in the most simplest case.

We start with reviewing the basic concepts of the fuzzy space described by the canonical tensor model.
The fuzzy space is a notion of space defined not by coordinates but by the algebra of linearly independent functions on the space;
the product of such functions, $f_a  (a=1,\cdots,N)$, is characterised  by a rank-three tensor:
\begin{equation}
f_a \star f_b = C_{ab}{}^c f_c. \label{ffcf}
\end{equation} 
To make contact with the canonical tensor model, 
we further impose two requirements \cite{Sasakura:2011ma}: 
\begin{enumerate}
\item reality conditions:
\begin{equation}
f^{\ast}_a = f_a, \ \ \ \left( f_a \star f_b \right)^{\ast} = f_b \star f_a,
\end{equation}
where $\ast$ stands for a complex conjugation;
\item a trace-like property:
\begin{equation}
\langle f_a | f_b \star f_c \rangle = \langle f_a \star f_b | f_c \rangle = \langle f_c \star f_a | f_b \rangle,
\end{equation}
where the inner product $\langle f_a | f_b \rangle$ has been chosen to be real, symmetric and bilinear.
\end{enumerate}
Since there exists a real linear transformation of $f_a$ which does not spoil the two requirements above,
without loss of generality one can choose the inner product as follows:
\begin{equation}
\langle f_a | f_b \rangle = \delta_{ab}, \label{delta}
\end{equation} 
if the inner product was set to be positive-definite as an initial condition.
Using (\ref{delta}), the degrees of freedom of the fuzzy space can be solely expressed by the rank-three tensor:
\begin{equation}
C_{abc} = \langle f_a \star f_b | f_c \rangle = C_{ab}{}^{d}\langle f_d | f_c \rangle.
\end{equation}
Since (\ref{delta}) is invariant under the orthogonal group transformation $O(N)$, 
the rank-three tensor $C_{abc}$ has the following kinematical symmetry:
\begin{equation}
C'_{abc} = J_a{}^d J_b{}^e J_c{}^f C_{def}, \ \ \ J \in O(N). \label{ortho}
\end{equation}
The two requirements on the function $f_a$ are translated into the generalised Hermiticity condition of $C_{abc}$:
\begin{equation}
C_{abc} = C_{bca} = C_{cab} = C^{\ast}_{bac} = C^{\ast}_{acb} = C^{\ast}_{cba}. \label{hermite}
\end{equation}

To make contact with general relativity, the time evolution of the fuzzy space (in other words, the rank-three tensor) is presumably generated by a ``local'' generator 
which somehow corresponds to the Hamiltonian constraint in general relativity. 
In addition, the system ought to have the invariance 
under the orthogonal group transformation (\ref{ortho}), which can be expected to correspond to the spatial diffeomorphism in general relativity.
Therefore, it is reasonable to define the Hamiltonian in such a way that the system becomes a constrained system
with the generators of the time evolution and the orthogonal transformation as first-class constraints. 
In this way, the total Hamiltonian of the canonical tensor model can be given as
\begin{equation}
H = N_a \mathcal{H}_a + N_{[ab]} \mathcal{J}_{[ab]} + N \mathcal{D}, \label{totalh}
\end{equation}
where $a,b,c=1,\cdots,N$; $[ab]$ denotes that $a$ and $b$ are anti-symmetric; $N_a$, $N_{[ab]}$ and $N$ are Lagrange multipliers;
\begin{equation}
\mathcal{H}_a =  P_{a(bc)} P_{bde} M_{cde}; \label{h}
\end{equation}
\begin{equation}
\mathcal{J}_{[ab]} = \frac{1}{2} \left( P_{acd}M_{bcd} - P_{bcd} M_{acd} \right); \label{j}
\end{equation}
\begin{equation}
\mathcal{D} = - \frac{1}{3}M_{abc}P_{abc}; \label{d}
\end{equation}
$P_{a(bc)}=\frac{1}{2}(P_{abc}+P_{acb})$.
As a convention, indices appearing repeatedly are summed from $1$ to $N$. 
Here $\mathcal{H}_a$ and $\mathcal{J}_{[ab]}$ are the generators of the time evolution and the orthogonal
group transformation, respectively; 
additionally $\mathcal{D}$, the generator of the scale transformation, has been introduced in order 
to regulate divergent behaviors of dynamics \cite{Sasakura:2013gxg}.
The rank-three tensors satisfy the following Poisson bracket:
\begin{equation}
\{ M_{abc}, P_{def} \} = \delta_{ad} \delta_{be} \delta_{cf} + (\text{cycric permutations of $(d,e,f)$});
\end{equation}
the other brackets vanish.
In fact, the form of the time-evolution generator $\mathcal{H}_a$ can be uniquely fixed,
if one imposes several reasonable assumptions, 
\textit{i.e.}, $(1)$ closed algebra, $(2)$ cubic terms at most, 
$(3)$ invariance under the time-reversal symmetry and $(5)$ connectivity \cite{Sasakura:2012fb}. 
The constrains, $\mathcal{H}_a$, $\mathcal{J}_{[ab]}$ and $\mathcal{D}$, are first-class: they form 
a first-class constraint algebra:
\begin{align}
&\{ H(T^{1}), H(T^2) \} = J([ \tilde{T}^1 , \tilde{T}^2 ]), 
\label{eq:dirachh}\\
& \{ J(V), H(T) \} = H(VT),
\label{eq:diracjh} \\
& \{ J(V^1), J(V^2)  \} = J([ V^1 , V^2 ]), 
\label{eq:diracjj}\\
&\{ \mathcal{D}, H(T) \} = H(T), 
\label{eq:diracdh}\\
&\{ \mathcal{D}, J(V) \} =0,
\label{eq:diracdj}
\end{align}
where $H(T) = T_a \mathcal{H}_a$, $J(V) =V_{[ab]}\mathcal{J}_{[ab]}$ and 
$\tilde{T}_{ab} = P_{(ab)c}T_c$; $[\ ,\ ]$ denotes the matrix commutator.
It has been pointed out \cite{Sasakura:2011sq} 
that this algebra has a close relationship with the Dirac algebra of general relativity 
\cite{Arnowitt:1960es,Dirac:1958sq,Dirac:1958sc,DeWitt:1967yk,Hojman:1976vp}.

From now we will examine a simple version of the canonical tensor model called \textit{minimal canonical tensor model} \cite{Sasakura:2013gxg}.
In this minimal model, the rank-three tensors, $M_{abc}$ and $P_{abc}$, are not Hermitian in the sense of (\ref{hermite}) but totally symmetric tensors.
In that case, one can consistently add a ``cosmological constant'' term $\lambda M_{abb}$ to $\mathcal{H}_a$,
if the constraint $\mathcal{D}$ is ignored, as was shown in the first part of \cite{Sasakura:2012fb};
the Hamiltonian becomes
\begin{equation}
H = N_a \mathcal{H}_a + N_{[ab]}\mathcal{J}_{[ab]}, \label{totalminih}
\end{equation}
where $\mathcal{J}_{[ab]}$ is the same as (\ref{j}), while $\mathcal{H}_a$ is changed to 
\begin{equation}
\mathcal{H}_a = P_{abc}P_{bde}M_{cde} - \lambda M_{abb}. \label{minih}
\end{equation}
The constraint algebra still takes the form given by (\ref{eq:dirachh}), (\ref{eq:diracjh}) and (\ref{eq:diracjj}),
while the part containing $\mathcal{D}$, (\ref{eq:diracdh}) and (\ref{eq:diracdj}), are discarded in this setting.

In order to extract some information of geometry from the rank-three tensors,
let us consider the minimal model with $N=1$: 
when $N=1$, the whole fuzzy space at some time slice can be described by a single function $f_1$.  
Here we rewrite the ingredients in (\ref{totalminih}) as follows: 
\begin{equation}
L \equiv \frac{1}{3} M_{111}, \ \ \ \Pi \equiv P_{111}, \ \ \ N \equiv 3 N_1, \ \ \  \Lambda \equiv  \lambda. \label{dictionary}
\end{equation}
By this convention, the Hamiltonian (\ref{h}) becomes
\begin{equation}
H = N \left( L \Pi^2 - \Lambda L \right), 
\end{equation}
with the Poisson bracket,
\begin{equation}
\{ L, \Pi \} =1.
\end{equation}
By the standard Legendre transformation, the corresponding action turns out to be
\begin{equation}
S_{\text{CT}}(L,N) = \int \text{d} t \left( \frac{\dot{L} (t)^2}{4N(t)L(t)} + \Lambda N(t) L(t)  \right), \label{stensor}
\end{equation}
where $\dot{L}$ denotes the time derivative of $L$.
What we will do in the next section is 
to compare this action with a minisuperspace action of general relativity in $d+1$ dimensions.

\section{Minisuperspace in general relativity}\label{mini}
In this section, for the purpose of interperting the variable $L$ in the action (\ref{stensor}) based on some known language, 
we consider $(d+1)$-dimensional general relativity in a minisuperspace. 
To begin with, remember the Einstein-Hilbert action with a cosmological constant $\Lambda$ in $d+1$ dimensions ($d>1$):
\be
S_{\text{EH}} = \frac{1}{16 \pi G_N} \int \text{d}^{d+1} x \left( R^{(d+1)} - 2 \Lambda \right),
\ee
where $G_N$ and $R^{(d+1)}$ are the Newton constant and the Ricci scalar in $d+1$ dimensional space-time, respectively. 
It has been given that space-time can be decomposed into space and time 
without breaking symmetry by Arnowitt, Deser and Misner (ADM) \cite{Arnowitt:1960es}:
\be
\text{d}s^2 = -N^2 \text{d}t^2 + h_{ij} (N^i\text{d}t + \text{d}x^i)(N^j\text{d}t + \text{d}x^j),
\ee       
where $N$, $N^i$ and $h_{ij}$ are a lapse function, a shift vector and a spatial metric, respectively;
the Latin indices run from $1$ to $d$. 
By applying the ADM decomposition, 
the Einstein-Hilbert action with the cosmological constant can be rewritten in the following form up to 
total derivative terms:
\be
S_{\text{EH}} =  \frac{1}{16 \pi G_N} \int  \text{d}t \text{d}^dx \sqrt{h} N \left( K_{ij}K^{ij} - K^2 + R^{(d)} - 2 \Lambda  \right), \label{adm}
\ee 
where $h$ is the determinant of the spatial metric; $K_{ij}$ is the extrinsic curvature defined as
\be
K_{ij} = \frac{1}{2N} ( \dot{h}_{ij} - \nabla_{i}N_{j} - \nabla_{j}N_{i} );
\ee
$K$ is a trace of the extrinsic curvature. Here $\nabla_i$ denotes the covariant derivative associated with the spatial metric.

Then we consider the following minisuperspace ansatz:
\be
N=N(t), \ \ \ N_i =0, \ \ \ h_{ij} = a(t)^2 \delta_{ij}, \label{cf}
\ee
where $a(t)$ is a scale factor of the Universe. Plugging the metric ansatz (\ref{cf}), one obtains the effective action:
\be
S_{\text{EH}}(a,N) = \frac{V}{16\pi G_N}  \int \text{d}t \  a^d \left(  d (1-d )  \frac{\dot{a}^2}{N a^2} - 2 \Lambda N     \right),  \label{eff}
\ee
where
\be
V = \int \text{d}^d x.
\ee
When the effective action (\ref{eff}) is written based on a quantity invariant under spatial diffeomorphism,
\be
L(t) = \int \text{d}^d x \sqrt{h(t)} = V a^d(t),
\label{eq:landarel}
\ee
one recovers (\ref{stensor}) up to some redefinitions of coupling constants: 
\be
S_{\text{EH}}(L,N) = \frac{\alpha}{16 \pi G_N} \int \text{d}t \left( \frac{\dot{L}(t)^2}{4N(t) L(t)} + \frac{d \Lambda}{2(d-1 )} N(t) L(t)   \right), \label{effl}
\ee
where
\be
\alpha = \frac{4 (1 - d )}{d}.
\ee
Note that the effective action (\ref{effl}) is universal: 
the effect of the dimensionality appears merely as the redefinition of coupling constants.  
As is clear from the ansatz we made (\ref{cf}), 
even if one adds higher spatial derivative terms by breaking Lorentz symmetry explicitly, 
the action still remains the same up to some redefinitions of coupling constants. 
For instance, in the case of Ho\v{r}ava-Lifshitz gravity \cite{Horava:2009uw}, 
one adds spatial derivative terms in such a way that the unitarity is preserved but the full space-time symmetry is broken down to the so-called \textit{foliation-preserving diffeomorphism}:
\be
S_{\text{HL}} =  \frac{1}{\kappa} \int  \text{d}t \text{d}^dx \sqrt{h} N \left( K_{ij}K^{ij} - \lambda K^2 + \gamma R^{(d)} - 2 \Lambda + \eta b^i b_i + \dots  \right), \label{hor}
\ee   
where $\kappa$, $\lambda$ $\gamma$, $\Lambda$ and $\eta$ are coupling constants; $b_i$ is a $d$-dimensional vector field \cite{Blas:2009qj}:
\be
b_i = \frac{\partial_i N}{N}.
\ee
The dots in (\ref{hor}) mean higher spatial derivative terms.
When taking the ansatz (\ref{cf}), one obtains
\be
S_{\text{HL}}(L,N) = \frac{\alpha}{\kappa} \int \text{d}t \left( \frac{\dot{L}(t)^2}{4N(t) L(t)} + \frac{d \Lambda}{2(d \lambda -1)} N(t) L(t)   \right), \label{effll}
\ee
where
\be
\alpha = \frac{4 (1 - d \lambda)}{d}.
\ee
Furthermore, the same form of the effective action has been obtained in the ($1+1$)-dimensional setup of Causal Dynamical Triangulations (CDT) 
\cite{Ambjorn:1998xu}
and the projectable Ho\v{r}ava-Lifshitz gravity \cite{Ambjorn:2013joa} without taking the minisuperspace ansatz like (\ref{cf}).
\footnote{
It has been shown that they are in the same universality class \cite{Ambjorn:2013joa}.}

\section{Summary and discussions}\label{dis}
In Section \ref{minimal}, we firstly have examined the minimal canonical tensor model with $N=1$ 
and obtained the effective action (\ref{stensor})
described by two functions of time, $L(t)$ and $N(t)$. 
Secondly in Section \ref{mini}, we have confirmed that the effective action (\ref{stensor}) coincides with the minisuperspace action of general relativity in arbitrary dimensions (\ref{eff}).

Let us closely look at the coincidence. 
In the canonical tensor model, we have set all the tensor slots to $1$; 
as a consequence, the generator of the orthogonal group transformation, $\mathcal{J}_{[ab]}$, vanishes.
Therefore, it can be considered that the spatial ``diffeomorphism'' (orthogonal group transformation) 
is gauged via the manipulation,
which is consistent with philosophy of the minisuperspace of general relativity.   


\section*{Acknowledgement}
We are very grateful to Bernard Raffaelli for useful discussions.
Y.S. thanks the Yukawa Institute for Theoretical Physics, 
where part of this work was done, for their warm hospitality.


\begin{thebibliography}{40}


\bibitem{Maldacena:1997re}
  J.~M.~Maldacena,
  ``The Large N limit of superconformal field theories and supergravity,''
  Adv.\ Theor.\ Math.\ Phys.\  {\bf 2} (1998) 231
  [hep-th/9711200].



\bibitem{Bekenstein:1981}
J.~D.~Bekenstein,
``Universal upper bound on the entropy-to-energy ratio for bounded systems,''
Phys.\ Rev.\ D {\bf 23} (1981) 287.

\bibitem{Bekenstein:2004sh}
  J.~D.~Bekenstein,
  ``How does the entropy / information bound work?,''
  Found.\ Phys.\  {\bf 35} (2005) 1805
  [quant-ph/0404042].



\bibitem{Sasakura:2011sq}
  N.~Sasakura,
  ``Canonical tensor models with local time,''
  Int.\ J.\ Mod.\ Phys.\ A {\bf 27} (2012) 1250020
  [arXiv:1111.2790 [hep-th]].

\bibitem{Sasakura:2012fb}
  N.~Sasakura,
  ``Uniqueness of canonical tensor model with local time,''
  Int.\ J.\ Mod.\ Phys.\ A {\bf 27} (2012) 1250096
  [arXiv:1203.0421 [hep-th]].

\bibitem{Sasakura:2013gxg}
  N.~Sasakura,
  ``A canonical rank-three tensor model with a scaling constraint,''
  Int.\ J.\ Mod.\ Phys.\ A {\bf 28} (2013) 1
  [arXiv:1302.1656 [hep-th]].

\bibitem{Sasakura:2013wza}
  N.~Sasakura,
  ``Quantum canonical tensor model and an exact wave function,''
  Int.\ J.\ Mod.\ Phys.\ A {\bf 28} (2013) 1350111
  [arXiv:1305.6389 [hep-th]].



\bibitem{ambjorn}
  J.~Ambjorn, B.~Durhuus and T.~Jonsson,
  ``Three-Dimensional Simplicial Quantum Gravity And Generalized Matrix
  Models,''
  Mod.\ Phys.\ Lett.\ A {\bf 6}, 1133 (1991).
 
\bibitem{sasakura}
N.~Sasakura,
``Tensor model for gravity and orientability of manifold,"
Mod.\ Phys.\ Lett.\ A {\bf 6} (1991) 2613.
  
  \bibitem{godfrey}
  N.~Godfrey and M.~Gross,
  ``Simplicial Quantum Gravity In More Than Two-Dimensions,''
  Phys.\ Rev.\ D {\bf 43}, 1749 (1991).


\bibitem{DePietri:2000ii}
  R.~De Pietri and C.~Petronio,
  ``Feynman diagrams of generalized matrix models 
      and the associated manifolds in dimension 4,''
  J.\ Math.\ Phys.\  {\bf 41} (2000) 6671
  [gr-qc/0004045].




\bibitem{Gurau:2009tw}
  R.~Gurau,
  ``Colored Group Field Theory,''
  Commun.\ Math.\ Phys.\  {\bf 304} (2011) 69
  [arXiv:0907.2582 [hep-th]].
  
\bibitem{Gurau:2011xp} 
  R.~Gurau and J.~P.~Ryan,
  ``Colored Tensor Models - a review,''
  SIGMA {\bf 8}, 020 (2012)
  [arXiv:1109.4812 [hep-th]].


\bibitem{Boulatov:1992vp}
  D.~V.~Boulatov,
  ``A Model of three-dimensional lattice gravity,''
  Mod.\ Phys.\ Lett.\ A {\bf 7} (1992) 1629
  [hep-th/9202074].

\bibitem{Ooguri:1992eb}
  H.~Ooguri,
  ``Topological lattice models in four-dimensions,''
  Mod.\ Phys.\ Lett.\ A {\bf 7} (1992) 2799
  [hep-th/9205090].

\bibitem{DePietri:1999bx}
  R.~De Pietri, L.~Freidel, K.~Krasnov and C.~Rovelli,
  ``Barrett-Crane model from a Boulatov-Ooguri field theory over a homogeneous space,''
  Nucl.\ Phys.\ B {\bf 574} (2000) 785
  [hep-th/9907154].

\bibitem{Freidel:2005qe}
  L.~Freidel,
  ``Group field theory: An Overview,''
  Int.\ J.\ Theor.\ Phys.\  {\bf 44} (2005) 1769
  [hep-th/0505016].

\bibitem{Oriti:2011jm}
  D.~Oriti,
  ``The microscopic dynamics of quantum space as a group field theory,''
  arXiv:1110.5606 [hep-th].





\bibitem{Sasakura:2011ma}
  N.~Sasakura,
  ``Tensor models and 3-ary algebras,''
  J.\ Math.\ Phys.\  {\bf 52} (2011) 103510
  [arXiv:1104.1463 [hep-th]].





\bibitem{Arnowitt:1960es}
  R.~L.~Arnowitt, S.~Deser and C.~W.~Misner,
  ``Canonical variables for general relativity,''
  Phys.\ Rev.\  {\bf 117}, 1595 (1960);
  R.~L.~Arnowitt, S.~Deser and C.~W.~Misner,
  ``The Dynamics of general relativity,''
  arXiv:gr-qc/0405109.
  

  
  
\bibitem{Dirac:1958sq} 
  P.~A.~M.~Dirac,
  ``Generalized Hamiltonian dynamics,''
  Proc.\ Roy.\ Soc.\ Lond.\ A {\bf 246}, 326 (1958).

\bibitem{Dirac:1958sc} 
  P.~A.~M.~Dirac,
  ``The Theory of gravitation in Hamiltonian form,''
  Proc.\ Roy.\ Soc.\ Lond.\ A {\bf 246}, 333 (1958).
  
\bibitem{DeWitt:1967yk} 
  B.~S.~DeWitt,
  ``Quantum Theory of Gravity. 1. The Canonical Theory,''  Phys.\ Rev.\  {\bf 160}, 1113 (1967).  

\bibitem{Hojman:1976vp} 
  S.~A.~Hojman, K.~Kuchar and C.~Teitelboim,
  ``Geometrodynamics Regained,''  Annals Phys.\  {\bf 96}, 88 (1976).  

\bibitem{Horava:2009uw}
  P.~Horava,
  ``Quantum Gravity at a Lifshitz Point,''
  Phys.\ Rev.\ D {\bf 79} (2009) 084008
  [arXiv:0901.3775 [hep-th]].

\bibitem{Blas:2009qj}
  D.~Blas, O.~Pujolas and S.~Sibiryakov,
  ``Consistent Extension of Horava Gravity,''
  Phys.\ Rev.\ Lett.\  {\bf 104} (2010) 181302
  [arXiv:0909.3525 [hep-th]].



  
\bibitem{Ambjorn:1998xu}
  J.~Ambjorn and R.~Loll,
  ``Nonperturbative Lorentzian quantum gravity, causality and topology change,''
  Nucl.\ Phys.\ B {\bf 536} (1998) 407
  [hep-th/9805108].



  
\bibitem{Ambjorn:2013joa}
  J.~Ambjorn, L.~Glaser, Y.~Sato and Y.~Watabiki,
  ``2d CDT is 2d Horava-Lifshitz quantum gravity,''
  Phys.\ Lett.\ B {\bf 722} (2013) 172
  [arXiv:1302.6359 [hep-th]].
  

 
  



\end{thebibliography}
\end{document}